\documentstyle[pre,multicol,aps]{revtex}

\begin{document}

\input{epsf.tex}
\title{Multistep cascading and fourth-harmonic generation}

\author{Andrey A. Sukhorukov, Tristram J. Alexander, and Yuri S. Kivshar}
\address{Australian Photonics Cooperative Research Centre, Research School of Physical Sciences and Engineering\\
Optical Sciences Centre, 
Australian National University, Canberra ACT 0200, Australia}

\author{Solomon M. Saltiel}
\address{Quantum Electronics Department, Faculty of Physics, 
     University of Sofia, Sofia 1164, Bulgaria}

\maketitle

\begin{abstract}
We apply the concept of multistep cascading to the problem of fourth-harmonic generation in a single quadratic crystal. We analyze a new model of parametric wave mixing and describe its stationary solutions for two- and three-color plane waves and spatial solitons. Some applications to the optical frequency division as well as the realization of the double-phase-matching processes in engineered QPM structures with phase reversal sequences are also discussed.
\end{abstract}

\begin{multicols}{2}
\narrowtext

Cascading effects in optical materials with quadratic (second-order or $\chi^{(2)}$) nonlinear response provide an efficient way to lower the critical power of all-optical switching devices~\cite{stegeman}. The concept of {\em multistep cascading}~\cite{concept} brings new ideas into this field, leading to the possibility of an enhanced nonlinearity-induced phase shift and generation of multicolor parametric spatial solitons. In particular, multistep cascading can be achieved by two nearly phase matched second-order nonlinear processes, second-harmonic generation (SHG) and sum-frequency mixing (SFM), involving the third-harmonic wave~\cite{solomon,solomon_soliton}. In this Letter, we extend the concept of multistep cascading to nonlinear effects of the fourth order and the fourth-harmonic generation (FHG) in a single noncentrosymmetric crystal. In particular, we analyze a new model of multistep cascading that involves the FHG process, and describe its stationary solutions for {\em normal modes}~--- plane waves and spatial solitons. Our study provides the first systematic analysis of the problem of FHG via a pure cascade process,  observed experimentally more than 25 years ago \cite{akh} and later studied in a cascading limit only~\cite{FHG}.

We consider the FHG via two second-order parametric processes: $\omega + \omega = 2 \omega$ and $2 \omega + 2 \omega = 4 \omega$, where $\omega$ is the frequency of the fundamental wave. In the approximation of slowly varying envelopes with the assumption of zero absorption of all interacting waves, we obtain
\[
  \begin{array}{l} {\displaystyle 
    \frac{\partial A}{\partial z} 
    = \frac{i}{2 k_1} \frac{\partial^{2} A}{\partial x^{2}} 
    + i \gamma_{1} A^{\ast} S e^{-i \Delta k_1 z} ,
  } \\*[9pt] {\displaystyle 
    \frac{\partial S}{\partial z} 
    = \frac{i}{2 k_2} \frac{\partial^{2} S}{\partial x^{2}} 
    + i \gamma_1 A^2  e^{i \Delta k_1 z}
    + i \gamma_2 S^{\ast} T e^{-i \Delta k_2 z} ,
  } \\*[9pt] {\displaystyle 
    \frac{\partial T}{\partial z} 
    = \frac{i}{2 k_4} \frac{\partial^{2} T}{\partial x^{2}} 
    + i \gamma_2 S^2 e^{ i \Delta k_2 z} ,
  } \end{array}
\]
where $A$, $S$, and $T$ are the envelopes of the fundamental-frequency~(FF), \mbox{second-}~(SH) and fourth-harmonic~(FH) waves respectively, $\gamma_{1,2}$ are proportional to the elements of the second-order susceptibility tensor, and $\Delta k_1 = 2 k_1 - k_2$ and $\Delta k_2 = 2 k_2 - k_4$ are the corresponding wave-vector mismatch parameters. We introduce the normalized envelopes ($u$, $v$, $w$) according to the following relations:
$A(x,z) = (16 z_d \gamma_2)^{-1} u( {x}/{a}, {z}/{2 z_d} )
          \exp(- i \Delta k_1 z / 2)$, 
$S(x,z) = (8 z_d \gamma_2)^{-1} v( {x}/{a}, {z}/{2 z_d} )$, and
$T(x,z) = (4 z_d \gamma_2)^{-1} w( {x}/{a}, {z}/{2 z_d} )
          \exp(i \Delta k_2 z)$, 
where $a$ is the characteristic beam width, and $z_d = k_1 a^2$ is the diffraction length of the FF component. 
In order to describe a family of nonlinear modes characterized by the {\em propagation constant} $\lambda$, we look for solutions in the form
$u(x,z) \rightarrow \lambda U( x \sqrt{|\lambda|}, z |\lambda| ) 
                     e^{i \lambda z/4}$,
$v(x,z) \rightarrow \lambda V( x \sqrt{|\lambda|}, z |\lambda| ) 
                     e^{i \lambda z/2}$, and
$w(x,z) \rightarrow \lambda W( x \sqrt{|\lambda|}, z |\lambda| ) 
                     e^{i \lambda z}$,
and obtain the normalized equations:
\begin{equation} \label{eq:uvw}
 \begin{array}{l} {\displaystyle 
    i s  \frac{\partial U}{\partial z}
    + s \frac{\partial^2 U}{\partial x^2} 
    - \alpha_1 U 
    + \chi U^{\ast} V  = 0,
  } \\*[9pt] {\displaystyle 
    2 i s  \frac{\partial V}{\partial z}
    + s \frac{d^2 V}{d x^2} 
    - V 
    + V^{\ast} W 
    + \frac{\chi}{2} U^2  = 0,
  } \\*[9pt] {\displaystyle 
    4 i s  \frac{\partial W}{\partial z}
    + s \frac{d^2 W}{d x^{2}} 
    - \alpha W 
    + \frac{1}{2} V^2 = 0.
  } \end{array}
\end{equation}
Here $s = {\rm sign} (\lambda) = \pm 1$, $\chi = \gamma_1 / (4 \gamma_2)$
is a relative strength of two parametric processes,
and the normalized mismatches are defined as
$\alpha = 4 + {\beta}/{\lambda}$ and
$\alpha_1 = {1}/{4} + {\beta_1}/{\lambda}$,
where $\beta = 8 \Delta k_2 z_d$ and $\beta_1 = - \Delta k_1 z_d$.

First, we analyze the plane-wave solutions of Eq.~(\ref{eq:uvw}) which do not depend on $x$. In this case, the total intensity $I$ is conserved, and we present it in terms of the unscaled variables as $I = I_u + I_v + I_w$, where $I_u = |u|^2 / 4$, $I_v = |v|^2$, and $I_w = 4 |w|^2$. Solutions $\{U_0,V_0,W_0\}$, which do not depend on $z$, are {\em the so-called normal modes}. The simplest one-component FH mode $\{0,0,W_0\}$ exists at $\alpha=0$. It has a fixed phase velocity $\lambda = - \beta/4$ and an arbitrary amplitude, being unstable for $I_w > \beta^2/4$ due to a {\em parametric decay instability}.

\begin{figure}
\setlength{\epsfxsize}{9.0cm}
\vspace*{-35mm}
\centerline{\mbox{\epsffile{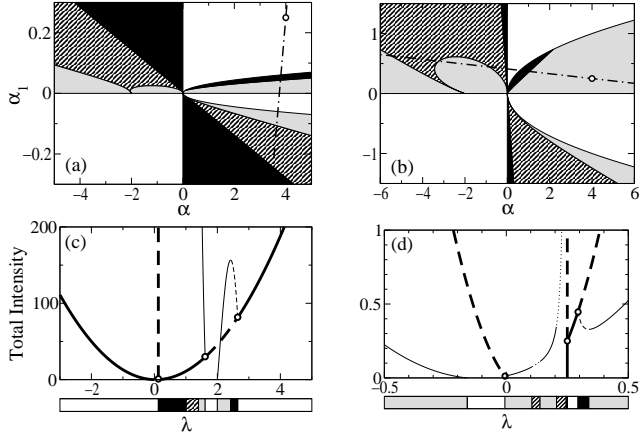}}}
\vspace*{-30mm}
\caption{ \label{fig:cw_pwr}
(a,b)~Existence and stability of three-mode plane waves for (a)~$\chi = \chi^{({\rm cr})}/4$ and (b)~$\chi = 4 \chi^{({\rm cr})}$. Light shading~--- stable, black~--- unstable, dark shading~--- oscillatory unstable, and blank~--- no solutions.  Open circles mark exact phase-matching, dash-dotted lines correspond to the lower plots. 
(c,d)~Intensity vs. $\lambda$ for (c)~$\beta = \beta_1 = -0.5$; (d)~$\beta = -1$ and $\beta_1 = 0.04$.  Thick dashed/solid vertical line~--- one-wave (FH) modes, thick curves~--- two-wave (SH + FH) modes, thin curves~--- three-wave modes.  Solid~--- stable, dashed~--- unstable, and dotted~--- oscillatory unstable modes.  Open circles mark the bifurcation points.
The legend beneath shows the stability of three-wave modes by using the same shadings as in the upper plots.} 
\end{figure}

Two-mode solution $\{0,\sqrt{2 \alpha },1\}$ describes a parametric coupling between SH and FH waves, and it exists for $\alpha>0$, bifurcating at $\alpha=0$ from the FH mode. Coupling of this two-mode plane wave to a FF wave can lead to its decay instability, provided
$|\alpha_1| < \alpha_1^{({\rm cr})} = \chi \sqrt{2\alpha}$. To understand the physical meaning of this inequality, we note that the family of solutions characterized by the propagation constant $\lambda$ corresponds to a straight line in the $(\alpha,\alpha_1)$--parameter space, see Figs.~\ref{fig:cw_pwr}(a,b). Moreover, all such lines include the point $(4,1/4)$ as the asymptotic limit for $|\lambda| \rightarrow +\infty$. This special point belongs to the instability region if the relative strength of the FF-SH interaction exceeds a {\em critical value}, i.e. for $\chi > \chi^{({\rm cr})} = 1/(8 \sqrt{2}) \simeq 0.088$.
However, for $\chi < \chi^{({\rm cr})}$ this {\em decay instability is suppressed for highly intense waves}, owing to a strong coupling with the FH field.

Finally, a three-mode solution, $V_0 = \alpha_1 / \chi$, $W_0 = V_0^2 / (2 \alpha)$, $U_0 = \sqrt{2 V_0 (1-W_0) / \chi}$, exists for (i)~$\alpha>0$ and $0<\alpha_1< \alpha_1^{({\rm cr})}$, (ii)~$\alpha>0$ and $\alpha_1< - \alpha_1^{({\rm cr})}$, and (iii)~$\alpha<0$ and $\alpha_1>0$.
In the limit $|\lambda| \rightarrow +\infty$, such three-wave modes are possible only for $\chi > \chi^{({\rm cr})}$. In the region~(i), stability properties of the three-wave solutions are determined by a simple criterion: the modes are stable if $\partial I / \partial |\lambda| >0$, and unstable, otherwise. For the parameter regions~(ii) and~(iii),  oscillatory instabilities are possible as well.
Existence and stability of all types of stationary plane-wave solutions of the model~(\ref{eq:uvw}) are summarized in Figs.~\ref{fig:cw_pwr}(a-d). 

\begin{figure}
\setlength{\epsfxsize}{9.0cm}
\vspace*{-50mm}
\centerline{\mbox{\epsffile{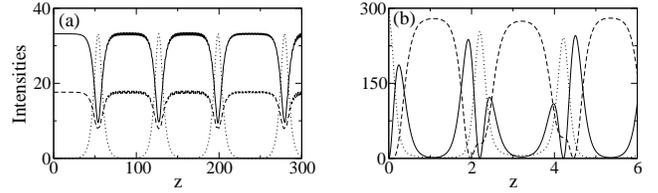}}}
\vspace*{-50mm}
\caption{ \label{fig:cw_nonst}
Dynamics of plane waves:
(a)~Instability of a two-wave mode [corresponding to $\lambda=2.1$ in Fig.~\ref{fig:cw_pwr}(c)],
(b)~Generation of higher harmonics from a FF input [parameters correspond to Fig.~\ref{fig:cw_pwr}(d)].
FF, SH, and FH components are shown by dotted, solid, and dashed curves, respectively. }
\end{figure}

In general, the system~(\ref{eq:uvw}) is nonintegrable and its dynamics are irregular. However, we find that in some cases a quasi-periodic energy exchange between the harmonics is possible. Figure~\ref{fig:cw_nonst}(a) shows one such case, when the intensities of unstable two-wave and stable three-wave stationary modes are close to each other, and an unstable two-wave mode periodically generates a FF component. Less regular dynamics are observed for other cases, such as for the generation of both SH and FH waves from an input FF wave [Fig.~\ref{fig:cw_nonst}(b)]. This example also illustrates the possibility of effective energy transfer to higher harmonics close to the double phase matching point.

Equations~(\ref{eq:uvw}) may have a different physical meaning provided the normalized amplitude $v$ stands for the mode of the fundamental frequency $\omega$. Then, Eqs.~(\ref{eq:uvw}) describe the optical frequency division by two (the field $u$) via parametric amplification and down-conversion (see, e.g.,~\cite{division}), provided both FF~($v$) and SH~($w$) fields are launched simultaneously at the input. Such a frequency division parametric process is in fact shown in Fig.~\ref{fig:cw_nonst}(a), where this time the generated $u$ wave corresponds to the frequency $\omega/2$, and is shown as dotted.

\begin{figure}
\vspace*{-35mm}
\setlength{\epsfxsize}{9.0cm}
\centerline{\mbox{\epsffile{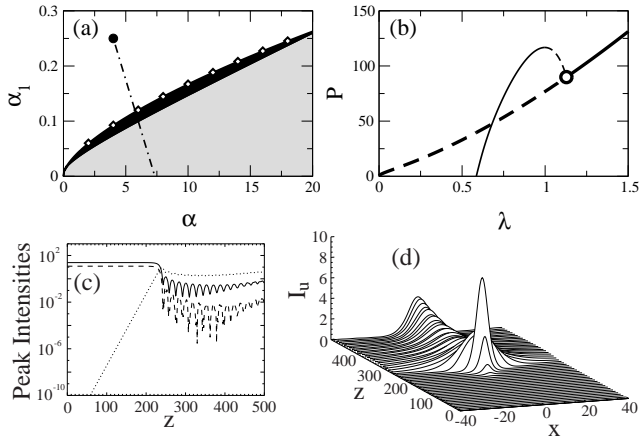}}}
\vspace*{-33mm}
\caption{ \label{fig:soliton}
(a)~Regions of existence and stability of three-mode parametric solitons
[shading is the same as in Fig.~\ref{fig:cw_pwr}(a)].
Open diamonds~--- an analytical approximation, dark circle~--- exact phase-matching point. 
The dash-dotted line corresponds to the solutions at $\beta = 2$ and $\beta_1 = -0.15$, for which the power vs. $\lambda$ dependences are shown in~(b): thick~--- two-wave (SH + FH), and thin~--- three-wave solitons; solid and dashed lines mark stable and unstable solutions, respectively. Open circle is the bifurcation point.
(c,d)~Development of a decay instability of a two-wave soliton corresponding to $\lambda = 1$ in~(b), and generation of a three-component soliton: 
(c)~FF, SH, and FH peak intensities vs. distance shown by dotted, solid, and dashed curves, respectively; 
(d)~evolution of the FF component.
For all the plots $\chi = \chi^{({\rm cr})}/2$.}
\end{figure}

We now look for spatially localized solutions of Eqs.~(\ref{eq:uvw}), {\em quadratic solitons}. First, we note that two-wave solitons consisting of the SH and FH components can be approximated as~\cite{concept,as}:
\begin{equation} \label{eq:solit_2wave}
  V_0(x) = {V_m}{{\rm sech}^{p}(x/p)}, \;\;\; 
  W_0(x) = {W_m}{{\rm sech}^{2}(x/p)},
\end{equation}
\[
  V_m^2 = \frac{\alpha W_m^2}{\left( W_m -1 \right)}, \;\; 
  \alpha = \frac{4 {\left( W_m-1 \right)}^3}{\left( 2-W_m \right)}, \;\; 
  p= \frac{1}{\left( W_m-1 \right)},
\]
where all parameters are functions of $\alpha$ only. Bright solitons either do not exist or are unstable being in resonance with linear waves~\cite{solomon_soliton} outside the parameter region $\alpha>0$ and $\alpha_1 > 0$, at $s=+1$. We find that in this region, similar to plane-wave modes, 
three-wave solitons exist for $0 < \alpha_1 < \alpha_1^{({\rm cr})}$, where the critical (cut-off) value $\alpha_1^{({\rm cr})}$ corresponds to a bifurcation from the two-wave solution~(\ref{eq:solit_2wave}). In order to find $\alpha_1^{({\rm cr})}$, we should solve the first equation of the system~(\ref{eq:uvw}) with the SH profile from Eq.~(\ref{eq:solit_2wave}) (see also \cite{concept}). However, such a linear eigenvalue problem has no exact analytical solution for arbitrary $p$, and thus we introduce an approximation $V_0(x) \simeq \widetilde{V}_0(x) = {V_m}{{\rm sech}^{2}(x/q)}$, requiring that the functions coincide at the amplitude level $V_m/2$, and then define the scaling parameter as
$q = p\; {{\rm cosh}^{-1}(2^{1/p})}/{ {\rm cosh}^{-1}(2^{1/2})}$.
Such an approximation adequately describes an effective soliton waveguide, and thus should provide overall good accuracy (except for some limiting cases). After solving the eigenvalue problem with the potential $\chi \widetilde{V}_0(x)$, we obtain an approximate expression for the {\em bifurcation points}:
$  \alpha_1^{({\rm cr})} 
    \simeq ( {\sqrt{1 + 4 V_m \chi q^2} - 1 - 2 n} )^2 
                  / (4 q^2)$, 
where $n$ is the order of the mode guided by the two-component parametric soliton waveguide~\cite{concept}. For a single-hump mode ($n=0$) the behavior of this cut-off is very similar to that of the plane waves. Indeed, in the cascading limit ($\alpha \gg 1$), we have $V_m \simeq 2 \sqrt{\alpha}$, and $\alpha_1^{({\rm cr})} \simeq 2 \chi \sqrt{\alpha}$, which differs by $\sqrt{2}$ from the corresponding result for plane waves. The critical value of $\chi$ for one-hump solitons can also be found from the approximate solution, $\chi^{({\rm cr})} \simeq 0.132$. We performed numerical simulations and found that the accuracy of our approximation is of the order of (and usually better than) 1\% in a wide range of parameters ($\chi > 10^{-2}$ and $\alpha > 10^{-2}$),
see Fig.~\ref{fig:soliton}(a).

Quite remarkably, for both positive $\alpha$ and $\alpha_1$ the stability properties of solitons [see  Figs.~\ref{fig:soliton}(a,b)] and plane waves [see Figs.~\ref{fig:cw_pwr}(a,c)] look similar. Specifically, stability of two- and three-component solitons is defined by the Vakhitov-Kolokolov criterion $\partial P/\partial \lambda > 0$, where $P=\int_{-\infty}^{+\infty} I d x$ is the soliton power, except for the region $\alpha_1 < \alpha_1^{({\rm cr})}$ where two-component solutions exhibit parametric decay instability. An example of such an instability is presented in Figs.~\ref{fig:soliton}(c,d), where an unstable two-wave soliton generates a stable three-wave state. Such instability-induced dynamics are very different from that of plane waves where, instead, quasi-periodic energy exchange is observed [see Fig.~\ref{fig:cw_nonst}(a)]. In the case of localized beams, diffraction leads to an effective power loss and convergence to a new (stable) state.

Similar to other models of multistep cascading~\cite{concept,solomon_soliton}, Eqs.~(\ref{eq:uvw}) possess various types of {\em exact analytical solutions}, which can be found at $\alpha = \alpha_1 = 1$ and $\chi > 1/\sqrt{2}$; $\alpha_1 = \alpha/4$ ($0<\alpha<1$) and $\chi = 1/(3\sqrt{2})$ or $\chi = [(3 \alpha) / (4 + 2 \alpha) ]^{1/2}$. Details will be presented elsewhere.

In order to observe experimentally the multistep cascading and multi-frequency parametric effects described above, we should satisfy the double-phase matching conditions. Using the conventional quasi-phase-matching~(QPM) technique~\cite{qpm} for FHG via a pure cascade process in LiTaO$_3$, we find that there exists only one wavelength ($2.45 {\rm \mu m}$), for which two parametric processes can be phase-matched simultaneously by the different orders $m$ of the QPM structure with the period $\Lambda_{\rm Q} = 34 {\rm \mu m}$. However, for the so-called phase-reversal QPM structures~\cite{reversal} characterized by two periods, the QPM period $\Lambda_{\rm Q}$ and the modulation period $\Lambda_{\rm ph}$ ($\Lambda_{\rm ph} > \Lambda_{\rm Q}$), double-phase matching is possible in a broad spectral range, provided the periods are selected to satisfy the conditions:
$   \Lambda_{\rm Q} = {4 \pi}
                      {\left| \Delta k_1 + \Delta k_2 \right|}^{-1}$,
$   \Lambda_{\rm ph} = {4 \pi m}
                      {\left| \Delta k_1 - \Delta k_2 \right|}^{-1}$,
where $m$ is the grating order. Thus, the engineered QPM structures suggested in~\cite{reversal} are more efficient than, e.g., the Fibonacci superlattices, and they can be used to achieve double-phase matching and to support different types of multistep cascading processes.

In conclusion, we have introduced a new model of multistep cascading that describes the fourth-harmonic generation via parametric wave mixing. We have analyzed the existence and stability of the stationary solutions of this model for normal modes~--- plane waves and spatial solitons. We have also discussed the possibility of double-phase-matching in engineered QPM structures with phase-reversal sequences.

\end{multicols}
\end{document}